\documentclass[onecolumn,useAMS,usenatbib,usegraphicx]{mn2e}

\newcommand{\apj}{ApJ}
\newcommand{\apjs}{ApJSS}
\newcommand{\mnras}{MNRAS}
\newcommand{\apjl}{ApJ Lett.} 
 
\newcommand{\aap}{A\&A}
\newcommand{\aj}{AJ}
\newcommand{\jcap}{JCAp}

\newcommand{\der}{{\rm d}}

\newcommand{\pk}{_{\rm pk}}

\newcommand{\tir}{\tilde r}

\newcommand{\xbra}{\lav\lav x\rav\rav}
\newcommand{\jpk}{_{\rm j}}
\newcommand{\opk}{_{\rm 1}}
\newcommand{\tpk}{_{\rm 2}}
\newcommand{\zpk}{_{\rm 3}}
\newcommand{\jp}{_{\rm p\,j}}

\newcommand{\R}{R_{\rm f}}

\newcommand{\ep}{e_{\rm p}}
\newcommand{\es}{e_{\rm s}}
\newcommand{\eps}{e_{\rm ps}}
\newcommand{\esp}{e_{\rm sp}}

\newcommand{\jj}{_{\rm j}}

\newcommand{\tang}{_{\rm t}}

\newcommand{\p}{_{\rm p}}
\newcommand{\rad}{_{\rm r}}

\newcommand{\modot}{M$_\odot$\ }

\newcommand{\e}{_{\rm e}}
\newcommand{\el}{^{\rm iso}}

\newcommand{\nbody}{{$N$}-body }
\newcommand{\sphc}{(r,\theta,\varphi)}
\newcommand{\beq}{\begin{equation}}
\newcommand{\eeq}{\end{equation}}
\newcommand{\beqa}{\begin{eqnarray}}
\newcommand{\eeqa}{\end{eqnarray}}
\newcommand{\lav}{\langle}
\newcommand{\rav}{\rangle}
\newcommand{\col}{_{\rm ta}}

\newcommand{\spot}{\lav\Phi\rav}
\newcommand{\srho}{\lav\rho\rav}

\begin{document}

\title[Halo Shape and Kinematics] {Theoretical dark matter halo
  kinematics and triaxial shape}

\author[Salvador-Sol\'e, Serra, Manrique \& Gonz\'alez-Casado]
{Eduard Salvador-Sol\'e$^1$\thanks{E-mail: e.salvador@ub.edu},
Sinue Serra$^1$, Alberto Manrique$^1$ \and{and Guillermo Gonz\'alez-Casado$^2$, }\\
$^1$Institut de Ci\`encies
del Cosmos,
Universitat de Barcelona (UB--IEEC), Mart{\'\i} i Franqu\`es 1,
E-08028 Barcelona, Spain\\
$^2$Dept.~Matem\`atica
Aplicada II, Centre de Recerca d'Aeron\`autica i de l'Espai
(UPC--IEEC), \\ Universitat Polit\`ecnica de Catalunya, E. Omega,
 Jordi Girona 1--3, E-08034 Barcelona, Spain}


\maketitle

\begin{abstract}
In a recent paper, \citet{Sea12} have derived the typical inner
structure of dark matter haloes from that of peaks in the initial
random Gaussian density field, determined by the power-spectrum of
density perturbations characterising the hierarchical cosmology under
consideration. In the present paper, we extend this formalism to the
typical kinematics and triaxial shape of haloes. Specifically, we
establish the link between such halo properties and the power-spectrum
of density perturbations through the typical shape of peaks. The
trends of the predicted typical halo shape, pseudo phase-space density
and anisotropy profiles are in good agreement with the results of
numerical simulations. Our model sheds light on the origin of the
power-law-like pseudo phase-space density profile for virialised
haloes.
\end{abstract}



\section{INTRODUCTION}\label{intro}

Virialised haloes in \nbody simulations of cold dark matter (CDM)
cosmologies show a wide variety of ellipsoidal shapes. On the
contrary, their structural and kinematic properties are remarkably
similar from one object to another. They are little sensitive not only
to the mass, redshift, environment and even specific cosmology, but
also to their individual shape. Only their scaling shows a mild
dependence on some of these properties. As shown by gravo-hydrodynanic
simulations, baryons introduce a larger scatter in the properties of
haloes at their central region. However, in the present paper, we will
concentrate on pure dark matter haloes and we will not deal with such
secondary baryonic effects.

The typical spherically averaged halo density profile, $\srho(r)$, is
well-fitted, down to about one hundredth the virial radius, by the
so-called NFW profile \citep{NFW97} as well as by the \citet{E65}
profile, which gives slightly better fits down to smaller radii
\citep{Navea04,M05,M06,St09,Navea10}. The velocity dispersion profile,
$\sigma(r)$, is reasonably well-fitted by the solution of the Jeans
equation for spherically symmetric isotropic systems with null value
at infinity \citep{CL96,M06}. More remarkably, \citet{TN01} showed
that the pseudo phase-space density profile is very nearly a pure
power-law,
\beq
\frac{\srho(r)}{\sigma^3(r)}= A r^{\nu}\,,\label{tn}
\eeq
with index $\nu\approx -1.875$ (see also
\citealt{As04,RTM04,Be07,FHGY07,VVKK09,Navea10}) and a similar
relation arises from the radial velocity dispersion component,
$\sigma\rad(r)$ \citep{DMc05}.  In equation (\ref{tn}), $\sigma(r)$ is
the velocity dispersion over the whole spherical shell with $r$, so
the variance coincides with the spherical average of the {\it local
  value} at points with $r$, $\sigma^2(r)=\lav\sigma_{\rm loc}^2
\rav(r)$. Finally, \citet{HM06} found that the velocity anisotropy
profile $\beta(r)$ behaves linearly with the logarithmic derivative of
the density,
\beq
\beta(r)=a \left(\frac{\der \ln \srho}{\der \ln r}+b\right)\,,
\label{hm}
\eeq
with $a$ and $b$ respectively equal to about $-0.2$ and $0.8$
(\citealt{HS06}; see also \citealt{Ludea11} for an alternative
expression), although with a substantial scatter this time.

The origin of all these trends is certainly related with the way dark
matter clusters. In hierarchical cosmologies, haloes grow through
continuous mergers with notably different dynamic effects according to
the relative mass of the captured and capturing objects. For this
reason, it is usually distinguished between major mergers, with a
dramatic effect each, and minor mergers, contributing together with
the capture of diffuse matter (if any) to the so-called accretion,
responsible of just a smooth secular evolution of the system. Some
authors have attempted to explain the typical halo density profile as
the result of repeated major (or intermediate) mergers
\citep{SW98,SSM98,Suea00,Dea03}.  Others have concentrated instead in
the effects of pure accretion (PA)
\citep{ARea98,NS99,DPea00,metal03,As04,Sea07}.  Both extreme scenarios
have also been investigated regarding the possible origin of the
pseudo phase-space density and velocity anisotropy profiles
\citep{HM06}. The PA scenario has received much support from the
results by \citet{WW09} showing that all typical halo trends are
already set in the first generation haloes formed by monolithic
collapse (i.e. no major merger; only accretion of diffuse matter) in
warm dark matter cosmologies.

Regarding the shape, CDM haloes are found to be triaxial ellipsoids,
with a trend towards prolate rather than oblate shapes (e.g.
\citealt{Fea88,dc91,Wa92,CL96,Sp04,All06,Ha07,Macea07,St09,VCea11}).
Inside each individual object, the typical minor to major axial ratio
takes a roughly uniform value of about $0.6$, with a slight trend to
an outward-decreasing triaxiality
\citep{Fea88,Bull02,JS02,Sp04,KE05,bs05,All06,Ha07,Be07,St09,VCea11}.
The main axis is preferentially aligned, at all scales, along with the
filament feeding the halo (e.g.
\citealt{LK99,BPYGT06,Patea06,Macea07,Ragea10,VCea11}). This indicates
that the memory of the preferred direction of major mergers and
accretion is not erased during virialisation \citep{VCea11} or,
equivalently, that the shape of virialised haloes depends on that of
their seeds. Moreover, as haloes are not supported by rotation but by
the local anisotropic velocity tensor, the fact that their triaxial
shape is related to the shape of protohaloes automatically implies
that their kinematics must also be related to it.

The seeds of haloes are believed to be peaks (secondary maxima) in the
primordial random Gaussian density field filtered at the scale of the
halo. The isodensity contours in the immediate vicinity of peaks are
triaxial \citep{Dor70} and rather prolate with a trend to become more
spherical for very high peaks (\citealt{BBKS}; hereafter BBKS). As
well-known, the monolithic collapse of non-spherical systems is highly
non-radial \citep{Z70}, giving rise to filaments and triaxial
virialised objects. Thus, it is natural to believe that the shape of
peaks is somehow translated into that of haloes, in agreement with the
above mentioned alignments. A few authors \citep{Lee05,RST11} have
tried to make the link between the shape of haloes and that of peaks
through the modelling of ellipsoidal collapse. Unfortunately, these
models do not account for the highly non-linear effects of
shell-crossing during virialisation, which play a crucial role in
setting the final properties of virialised haloes. On the other hand,
there is in the literature no attempt to relate the kinematics of
haloes with the shape of their seeds.

In a recent paper, \citeauthor{Sea12} (\citeyear{Sea12}; hereafter
SVMS) have shown that the kinematics of virialised haloes in
(bottom-up) hierarchical cosmologies with dissipationless
collisionless dark matter depends on their triaxial shape, contrarily
to their spherically averaged density profile which does not. This
allowed SVMS to infer, under the assumption of PA, the typical
spherically averaged density profile for haloes from that of peaks in
the primordial density field, determined by the power-spectrum of
density perturbations. Furthermore, SVMS showed that the density
profile for haloes having undergone major mergers is indistinguishable
from that of haloes grown by PA, so the model actually holds for all
haloes regardless of their individual aggregation history.

In the present paper, we extend the SVMS model to the kinematics and
triaxial shape of virialised objects. Under the PA assumption and
neglecting any possible rotation tidally induced by surrounding
matter, we derive the halo shape, velocity anisotropy profile and
pseudo phase-space density profile from the triaxial shape of peaks,
taking into account the full virialisation process. We first assume
the simple case of PA and then study the foreseeable effects of major
mergers. This allows us to establish the link between those typical
halo properties and the power-spectrum of density perturbations. The
theoretical predictions obtained when this formalism is applied to CDM
haloes are in good agreement with the results of numerical
simulations.

The paper is organised as follows. In Section \ref{axratio}, we derive
some general relations valid for triaxial systems, regardless of
whether they are in equilibrium or not. Assuming PA, these relations
are used, in Section \ref{eccentricity}, to make the link between the
triaxial shape of a virialised object formed by PA and that of its
seed. The typical velocity anisotropy and velocity dispersion profiles
for virialised objects are derived in Section \ref{anisotropy} from
their triaxial shape.  In Section \ref{haloes}, we apply the model to
CDM haloes. The origin of the power-law-like form of the pseudo
phase-space density profile and the effects of major mergers are
discussed in Section \ref{mm}. Our results are summarised in Section
\ref{summ}.  A package with the numerical codes used in the present
paper is publicly available from {\texttt
  {www.am.ub.es/$\sim$cosmo/haloes\&peaks.tgz}}.

\section{Mean Squared and Crossed Fluctuation Profiles}\label{axratio}

There are in the literature several ways to characterise the shape of
a triaxial system with semiaxes $a\ge b\ge c$: the axial ratios, the
ellipticity and prolateness (e.g. BBKS) and the triaxiality parameter
(e.g. \citealt{Fea91}), among others. In the present paper, we use
the primary and secondary eccentricities, respectively defined as
\beq
\ep=\left(1-\frac{c^2}{a^2}\right)^{1/2}~~~~~~~~~{\rm and}~~~~~~~~~~\es=\left(1-\frac{b^2}{a^2}\right)^{1/2}.
\label{primary}
\eeq
As the eccentricities vary, in general, over the system (we assume
from now on that the centre of symmetry remains the same at all
scales), we will deal with the eccentricity profiles $\ep(r)$ and
$\es(r)$, being $r$ the radius of a sphere associated with each
ellipsoid (see details below).

Numerical studies of virialised haloes usually work with spherically
averaged profiles. Of course, triaxial systems show non-null
departures $\delta\rho\sphc$ and $\delta\Phi\sphc$ of the local
density $\rho\sphc$ and gravitational potential $\Phi\sphc$ from their
respective spherical averages $\srho(r)$ and $\spot(r)$. Hence, the
(scaled) mean squared or crossed density and potential fluctuation
profiles, $\lav (\delta\rho/\srho)^2\rav(r)$, $\lav
(\delta\Phi/\spot)^2\rav(r)$ and $\lav (\delta\rho/\srho)
(\delta\Phi/\spot)\rav(r)$, also give a measure of the triaxial shape
of those systems. Below, we relate these profiles to the eccentricity
profiles defined above.

Take the Cartesian axis $z$ aligned along with the major axis of the
ellipsoidal isodensity contour $\rho\el(r)$, labelled by the radius
\beq
r=\left[\frac{1}{3}\left(a^2+b^2+c^2\right)\right]^{1/2}\,.
\label{R1}
\eeq
The local density at the point $\sphc$ then takes the form
\beq
\rho\sphc=\rho\el(r)\left[1-\frac{\ep^2(r)+\es^2(r)}{3}\right]
\left[\sin^2\theta \cos^2\phi+\frac{\sin^2\theta \sin^2\phi}{1-\esp^2(r)}+
\frac{\cos^2\theta}{1-\eps^2(r)}\right]\,,
\label{ellips}
\eeq
where $\eps$ stands for one of the two eccentricities and $\esp$ for the
other one\footnote{The specific values of $\eps$ and $\esp$ depend
  on the orientation of the $x$ and $y$ Cartesian axes relative to the
  minor and intermediate semiaxes.}. The spherically averaged density
at $r$ is then equal to
\beq
\srho(r)=\frac{\rho\el(r)}{3}\left[1-\frac{\ep^2(r)+\es^2(r)}{3}\right]G(r),
\label{norm}
\eeq
where we have defined the function
\beq 
G(r)=\left[1+\frac{1}{1-\es^2(r)}+\frac{1}{1-\ep^2(r)}\right]
=a^2(r)\left[\frac{1}{a^2(r)}+\frac{1}{b^2(r)}+\frac{1}{c^2(r)}\right]\,.
\eeq
Dividing equation (\ref{ellips}) by $\srho(r)$ and replacing
$\rho\el(r)$ by its value given in equation (\ref{norm}), we obtain
\beq
1+\frac{\delta\rho}{\srho}(r,\theta,\phi)=\frac{3}{G(r)}
\left[\sin^2\theta \cos^2\phi+\frac{\sin^2\theta \sin^2\phi}{1-\esp^2(r)}+
\frac{\cos^2\theta}{1-\eps^2(r)}\right]\,.
\label{ellips2}
\eeq
The mean squared density fluctuation over the sphere with radius $r$,
\beq
\left\lav\left(\frac{\delta\rho}{\srho}\right)^2\right\rav(r)=\frac{1}{4\pi}\int_0^\pi \der\theta \sin\theta\int_0^{2\pi} \der \phi\left(\frac{\delta\rho}{\srho}\right)^2\!\!(r,\theta,\phi)\,,
\label{rms}
\eeq
can then be readily obtained from $\delta\rho/\srho$ given by equation
(\ref{ellips2}). The result is
\beq
\left\lav\left(\frac{\delta\rho}{\srho}\right)^2\right\rav(r)= -\frac{2}{5}\left\{1-
\frac{3[(1-\ep^2)^2(1-\es^2)^2+(1-\ep^2)^2+(1-\es^2)^2]}
{[(1-\ep^2)(1-\es^2)+(1-\ep^2)+(1-\es^2)]^2}\right\}(r)\,.
\label{4th}
\eeq

The relations between the eccentricity profiles and the mean squared
potential and crossed density--potential fluctuation profiles follow
from the relation (\ref{4th}) and the relations between such fluctuation profiles
and the mean squared density fluctuation profile. To derive them we
need first to develop the member on the left of the Poisson equation,
\beq
\nabla^2\left[\spot\left(1+\frac{\delta\Phi}{\spot}\right)\right]=4\pi
G \srho\left(1+\frac{\delta\rho}{\srho}\right)\,.
\label{poisson}
\eeq 
Taking into account that, by the Gauss theorem, $\spot$ satisfies
the usual Poisson integral relation for spherically symmetric systems
\beq
\frac{\der\spot(r)}{\der r} = \frac{G M(r)}{r^2}\,,
\label{gauss}
\eeq
we arrive after some algebra at the exact relation
\beq
\frac{\delta\rho}{\srho}=\frac{\delta\Phi}{\spot}-2\xi(r)r
\,\frac{\partial}{\partial r}\left(\frac{\delta\Phi}{\spot}\right)-
\xi(r)\zeta(r)r^2\,\nabla^2\left(\frac{\delta\Phi}{\spot}\right)\,,
\label{Binney}
\eeq
where $\xi(r)=(\der \ln M/\der \ln r)^{-1}$ and $\zeta(r)=-(\der \ln
|\spot|/\der \ln r)^{-1}$. Equation (\ref{Binney}) shows that it is
possible to infer $\delta\Phi/\spot$ from $\delta\rho/\srho$ for the
appropriate boundary conditions. This requires, however, the full
characterisation of $\delta\rho\sphc$, which is, in general, a
complicate function of the spatial coordinates. For simplicity, we
will concentrate in ellipsoidal systems (i) {\it with non-rotating
  symmetry axes} and (ii) {\it with rms density fluctuation profile of
  the power-law form}, $\lav (\delta\rho/\srho)^2\rav^{1/2} = Q
r^\kappa$, or, equivalently, satisfying $\partial
[\delta\rho\sphc/\srho(r)]/\partial r = \kappa r^{-1}
\delta\rho\sphc/\srho(r)$.\footnote{Such an equivalence can be proven
  taking into account that, as $\lav \delta\rho/\srho\rav =0$, either
  the integral of $\delta\rho\sphc/\srho(r)$ over $\phi$ is an even
  function of $\theta$ or the integral of $\sin
  (\theta)\delta\rho\sphc/\srho(r)$ over $\theta$ is an odd function
  of $\phi$ and that the condition $\lav
  (\delta\rho/\srho)^2\rav^{1/2} = Q r^\kappa$ implies the contrary
  relation for $\{\partial [\delta\rho\sphc/\srho(r)]/\partial
  r\}-\kappa [\delta\rho\sphc/\srho(r)]/ r$.} Condition (i) is
essentially satisfied by simulated haloes (see
\citealt{JS02,Sp04,KE05,bs05,Ha07,VCea11}), while condition (ii) can
be seen as a reasonable approximation, at least within some finite
  radial range\footnote{In fact, an arbitrarily accurate approximation
    can be achieved by splitting the real log-log mean square density
    fluctuation in a series of concatenate linear functions so that
    such a condition would hold in each small segment.}.

In these conditions, multiplying equation (\ref{Binney}) by the scaled
density fluctuation, performing the spherical average and taking into
account the divergence theorem, we are led to
\beq
\left\lav \left(\frac{\delta\rho}{\srho}\right)^{\!2}\right\rav(r)=
K(r)\left\lav\frac{\delta\rho}{\srho}\frac{\delta\Phi}{\spot}\right\rav(r)
-\xi(r)\zeta(r)r^2\left\{\frac{2\left[1+(1-\kappa)\zeta(r)\right]}{\zeta(r)r}\frac{\der}{\der r}\left\lav\frac{\delta\rho}{\srho}\frac{\delta\Phi}{\spot}\right\rav
+\frac{\der^2 }{\der r^2}\left\lav
\frac{\delta\rho}{\srho}\frac{\delta\Phi}{\spot}\right\rav\right\},
\label{D1}
\eeq
being $K(r)=1+\kappa\xi(r)[2+(1-\kappa)\zeta(r)]$.  Equation
(\ref{D1}) is an ordinary differential equation that can be solved,
for trivial consistency boundary condition at $r=0$, for the mean
crossed density--potential fluctuation.  Similarly, multiplying
equation (\ref{Binney}) by the scaled potential fluctuation, the same
derivation as above leads to
\beq
\left\lav\frac{\delta\rho}{\srho}\frac{\delta\Phi}{\spot}\right\rav(r)=
\left\lav\left(\frac{\delta\Phi}{\spot}\right)^2\right\rav(r)
-\xi(r)\zeta(r)r^2\left\{\frac{2[1+\zeta(r)]}{\zeta(r)r}\frac{\der}{\der r}\left\lav\left(\frac{\delta\Phi}{\spot}\right)^2\right\rav+
\frac{\der^2}{\der r^2}\left\lav\left(\frac{\delta\Phi}{\spot}\right)^2\right\rav
-\left\lav\left(\frac{\partial}{\partial r}\frac{\delta\Phi}{\spot}\right)^2\right\rav
\right\}.
\label{5thbis}
\eeq
Equation (\ref{5thbis}) cannot yet be solved for the mean squared
potential fluctuation because of the presence of the unknown mean
squared radial derivative of the scaled potential (the last term on
the right). But this drawback can be bypassed by multiplying equation
(\ref{Binney}) by the partial derivative of the scaled
potential fluctuation and operating as usual. This leads to
\beq
\frac{\der}{\der r}\!\left\lav\frac{\delta\rho}{\srho}\frac{\delta\Phi}{\spot}\right\rav\!-\!\frac{\kappa}{r}\left\lav\frac{\delta\rho}{\srho}\frac{\delta\Phi}{\spot}\right\rav\!(r)=\frac{1}{2}\frac{\der}{\der r}\!\left\lav\!\left(\frac{\delta\Phi}{\spot}
\right)^{\!\!2}\right\rav
\!-\!\xi(r)\zeta(r)r\left\{\frac{2\left[1+\zeta(r)\right]}{\zeta(r)}\!\left\lav\!\left(\frac{\partial}{\partial r}\frac{\delta\Phi}{\spot}\right)^{\!\!2}\right\rav\!(r)\!+\!\frac{1}{2}
\frac{\der }{\der r}\!\left\lav\!\left(\frac{\partial}{\partial r}\frac{\delta\Phi}{\spot}\!\right)^{\!\!2}\right\rav\right\}\!.
\label{D2}
\eeq
Then, substituting in equation (\ref{D2}) the mean squared partial
derivative of the scaled potential given by equation
(\ref{5thbis}) and its radial derivative drawn from the
differentiation of the same equation, we arrive at
\beqa
\left\lav\frac{\delta\rho}{\srho}\frac{\delta\Phi}{\spot}\right\rav(r)+\frac{3\,r}{4\widetilde I(r)}\frac{\der}{\der r}
\left\lav\frac{\delta\rho}{\srho}\frac{\delta\Phi}{\spot}\right\rav=\frac{I(r)}{\widetilde I(r)}
\left\lav\left(\frac{\delta\Phi}{\spot}\right)^2\right\rav(r)
~~~~~~~~~~~~~~~~~~~~~~~~~~~~~~~~~~~~~~~~~~~~~~~~~~~~~~~~\nonumber\\
-\frac{3\xi(r)r}{4\widetilde I(r)}\left\{J(r)\frac{\der}{\der r}\!\left\lav\left(\frac{\delta\Phi}{\spot}\right)^{\!2}\right\rav\!+[1+\zeta(r)]r
\frac{\der^2}{\der r^2}\left\lav\left(\frac{\delta\Phi}{\spot}\right)^{\!2}\right\rav
+\frac{\zeta(r)r^2}{6}
\frac{\der^3}{\der r^3}\left\lav\left(\frac{\delta\Phi}{\spot}\right)^{\!2}\right\rav\right\},
\label{D3}
\eeqa
being
\beq
J(r)=\frac{1}{3}\left[4I(r)+3\zeta(r)-2\xi^{-1}(r)+\frac{\der\ln
\xi}{\der \ln r}+5\right]~,~~~~~~~~~~~
I(r)= 1+\zeta^{-1}(r)+\frac{1}{4}\frac{\der \ln}{\der\ln r}\left(\frac{\zeta}{\xi\,r^2}\right) 
\label{H}
\eeq
and $\widetilde I(r)=I(r)-\kappa/2$. Once again, equation (\ref{D3}) can be
solved, for trivial consistency boundary conditions at $r=0$, for the
mean squared potential fluctuation. Note that the mean
$\partial(\delta\rho/\srho)/\partial r$ $\delta \Phi/\spot$ profile
entering the Jeans equation for anisotropic spherically averaged
triaxial systems (eq.~[\ref{exJeq2}] below) can also be inferred from
the radial derivative of the mean crossed density-potential
fluctuation and the condition $\partial
[\delta\rho\sphc/\srho(r)]/\partial r = \kappa r^{-1}
\delta\rho\sphc/\srho(r)$ above.

One particular case of interest is that of homologous triaxial
systems, i.e. with $\kappa=0$. Equations (\ref{D1}), (\ref{5thbis})
and (\ref{D3}) then have the trivial solution,
\beq
\left\lav\frac{\delta\rho}{\srho}\frac{\delta\Phi}{\spot}\right\rav(r)=
\left\lav\left(\frac{\delta\Phi}{\spot}\right)^2\right\rav(r)=
\left\lav\left(\frac{\delta\rho}{\srho}\right)^2\right\rav(r)= Q\,.
\label{constant}
\eeq
That is, all squared or crossed fluctuation profiles are uniform and
equal to each other.

\begin{figure}
\vskip -0.7 cm
\centerline{\includegraphics[scale=1.]{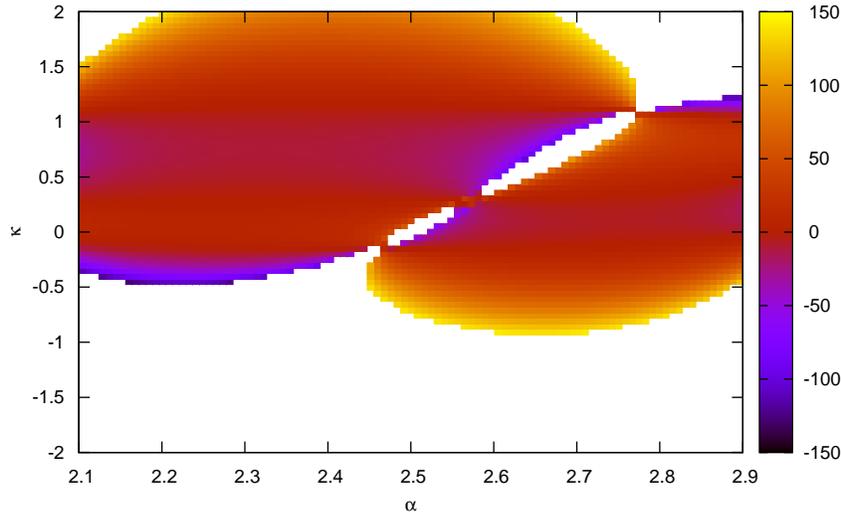}}
\caption{Predicted $Q/P$ ratio (see the colour bar for the
    different ranges of this quantity) between the rms density and potential
  fluctuation profiles in self-similar objects with $\kappa$ the power
  index of the rms density and potential fluctuations and $-\alpha$
  that of the spherically averaged density profile.}\label{f2}
\end{figure}

Another particular case of interest is that found in asymptotic
regimes, i.e. with all the profiles behaving as power-laws. Equations
(\ref{5thbis}), (\ref{D2}) and (\ref{D3}) then imply that both the rms
density and potential fluctuation profiles have the same form,
proportional to $r^{\kappa}$. This does not mean that the isodensity
and isopotential contours are equally non-spherical at any given
radius: the {\it positive} proportionality factors $Q$ and $P$ of the
rms density and potential fluctuation profiles, respectively, are
different in general. Specifically, for an object with power-law
spherically averaged density profile $\srho(r)\propto r^{-\alpha}$,
the ratio $Q/P$ depends on $\kappa$ and $\alpha$ in the following way,
\beq
\frac{Q}{P}=\left[1-\frac{(1-\alpha)(\kappa-1)\kappa}{(2-\alpha)(3-\alpha)}\right]\left(1-\frac{\frac{3\kappa}{2}\left\{1-\frac{(\kappa-1)}{(2-\alpha)}\left[\frac{1}{\kappa-1}-\frac{(\kappa-2)}{6\,\kappa\,(3-\alpha)}\right]\right\}}{5-2\alpha+\kappa/2}\right)\,,
\label{QP}
\eeq 
as inferred from equations (\ref{D1}) and (\ref{D3}).  Figure \ref{f2}
shows the ratio $Q/P$ that results for $\alpha$ and $\kappa$
respectively in the ranges $2 < \alpha < 3$ and $-2 <\kappa < 2$. The
function is well-behaved everywhere except in the line
$\kappa=4\alpha-10$ where it diverges by alternating positive and
nonphysical negative values on opposite sides of the line, in a series
of segments separated by a few well-behaved points. Far enough from
the strip, $Q/P$ tends to take values greater than one for any value
of $\kappa$.

Notice that, as $\rho\sphc$ is positive, the negative values of
$\delta\rho$ necessarily satisfy the condition
$|\delta\rho|<\srho$. Moreover, in rather prolate ellipsoidal systems
the solid angle where $\delta \rho$ is negative is less than the solid
angle where it is positive and, as the mean $\delta \rho$ is null, we
also necessarily have that the positive values of $\delta \rho$ are
less than $\srho$. Consequently, we have $|\delta\rho/\srho| < 1$, in
general. Then, provided the mass distribution is smooth enough, which
is guaranteed in triaxial virialised objects, equation (\ref{Binney})
also implies $|\delta\Phi/\spot| < 1$. The angular cross-correlation
of these two dimensionless quantities is even smaller, so any physical
quantity reporting to the triaxial system can be written as a series
expansion of the dimensionless fluctuation profiles measuring the
deviation from spherical symmetry. We will take advantage of this fact
in Sections \ref{eccentricity} and \ref{anisotropy}.

All the previous relations hold regardless of whether the triaxial
system is in equilibrium or not. Thus, they apply both to the
virialised object and its seed. From now on, all quantities referring
to the protoobject are denoted by subindex p.

\section{ECCENTRICITY PROFILES}\label{eccentricity}

We want to relate the eccentricities of a virialised triaxial object
with those of its seed. As there are two eccentricities, we need two
equations. These equations must obviously report on properties
involving the shape of these systems.

One of such properties is the volume of ellipsoidal isodensity
contours. The ratio between the volumes of the ellipsoids encompassing
a given mass in the final and initial systems must be equal, to
leading order in the deviation from spherical symmetry, to the ratio
between the volumes of the corresponding spheres encompassing
identical mass,
\beq
\frac{a(r)\,b(r)\,c(r)}{a\p(r\p)\,b\p(r\p)\,c\p(r\p)}=\frac{r^3}{r\p^3(r)}\,.
\label{emass}
\eeq
To write the member on the right of equation (\ref{emass}) we have
taken into account that the mass of an ellipsoidal isodensity contour
with semiaxes, $a$, $b$ and $c$, labelled by $R$
\beq 
M(R)=\int_0^\pi \der\theta \sin\theta\int_0^{2\pi} \der \phi \int_0^R \der r\,r^2\,\rho(r,\theta,\phi)\,,
\label{mass1}
\eeq
coincides, owing to the relations (\ref{ellips}) and (\ref{norm})
arising from the definition (\ref{R1}) of the labelling radius, with
the mass of the sphere with radius $R$,
\beq
M(R)=4\pi \int_0^{R} \der r\,r^2\srho(r)\,.
\label{mass}
\eeq
On the other hand, the function $r\p(r)$ in equation (\ref{emass}) is
the solution, for the boundary condition $r\p(0)=0$, of the
differential equation
\beq
\frac{\der r\p}{\der r}=\frac{r^2\srho(r)}{r\p^2\,\lav \rho\p\rav[r\p(r)]}\,,
\label{rpr}
\eeq
that follows from differentiation of equation (\ref{mass}) holding for
spheres with identical mass in the initial and final systems. For
simplicity in the notation, we omit from now on the explicit
dependence of $r\p$ on $r$.

We stress that equation (\ref{emass}) is only valid to leading order
in the deviations from spherical symmetry. This does not mean that the
volume of each ellipsoid is approximated by that of the corresponding
sphere. (By doing this, we would lose the information on the shape of
the system.)  What we approximate is the whole ratio between the
volumes of both ellipsoids by the ratio between the volumes of the
corresponding spheres. Were the axial ratios of the ellipsoid
conserved over the evolution of the system, the relation (\ref{emass})
would be exact. Actually, the axial ratios vary during
virialisation. But this variation is of higher order in the deviation
from spherical symmetry, so the ratio between the volumes of the two
ellipsoids is kept equal to that between the two spheres to leading
order.

Taking into account the relation (\ref{primary}), equation (\ref{emass}) leads to
\beq
\frac{(1-\ep^2)(1-\es^2)}{[1+(1-\ep^2)+(1-\es^2)]^3}(r)=
\frac{(1-\ep^2)(1-\es^2)}{[1+(1-\ep^2)+(1-\es^2)]^3}(r\p)\,.
\label{6th}
\eeq

A second property involving the triaxial shape of the system is the
energy of the sphere with radius $R$ encompassing a fixed mass $M$
(see SVMS),
\beq
E(R)=4\pi \int_0^R \der r\,r^2\srho(r) \left[\frac{\sigma^2(r)}{2}-\frac{GM(r)}{r}\right]+ 
2\pi \int_0^R \der r\,r^2\lav\delta\rho\,\delta\Phi\rav(r)\,.
\label{ener}
\eeq
This can be written in the form $E(R)={\cal E}(R)+\delta{\cal
  E}(R)$. The so-called ``spherical'' total energy,
\beq 
{\cal E}(R)= 4\pi \int_0^{R} \der
r\,r^2\srho(r)\left[\frac{s^2(r)}{2}-\frac{GM(r)}{r}\right]\,,
\label{energy}
\eeq
measures the total energy in the sphere with $M$, were the mass
distribution inside it spherically symmetric and endowed with the
velocity variance profile $s^2(r)$ that would result taking into
account the energy lost by shell-crossing during virialisation but
not the potential energy exchanged with the rest of the system owing
to its non-spherical mass distribution. In the protoobject,
$s^2\p(r\p)$ can be taken equal to the real velocity variance profile
$\sigma\p^2(r)$, but in the virialised object $s^2(r)$ differs from
$\sigma^2(r)$. Nonetheless, the profile $s^2(r)$
can be obtained, by differentiation of equation (\ref{energy}),
\beq 
s^2(r)=2\left[\frac{\der {\cal E}/\der
r} {\der M/\der r}+ \frac{GM(r)}{r}\right],
\label{sig2}
\eeq 
from the mass and spherical energy profiles, $M(r)$ and ${\cal E}(r)$,
respectively, derived in SVMS. The result was
\beq 
r(M)= -\frac{3}{10}\,\frac{GM^2}{{\cal E}\p(M)}\,
\label{vir0}
\eeq
and 
\beq 
{\cal E}(R)=-R\int_0^R \der r\left[\frac{GM(r)}{r^2}\frac{\der M}{\der r}+\frac{{\cal W}(r)}{2r^2}\right]\,,
\label{diff}
\eeq
where we have defined the ``spherical'' potential energy, 
\beq
{\cal W}(R)=-4\pi\int_0^R \der r\,r^2\srho(r)\,\frac{GM(r)}{r}\,.
\label{spot2}
\eeq
This leads to the following explicit expression
\beq 
s^2(r)=-\frac{1}{4\pi r^2\srho(r)}\left\{\frac{{\cal
    W}(r)}{r}+ \int_0^r \der\tir \left[8\pi \srho(\tir)GM(\tir)+\frac{{\cal
      W}(\tir)}{\tir^2}\right]\right\}\,,
\eeq 
showing that $s^2(r)$ can be calculated from the spherically averaged
density profile $\srho(r)$. As this latter profile is independent of
the shape of the system, so must the function ${\cal E}(r)$ (see
eq.~[\ref{energy}]). Consequently, all the effects of triaxiality in
$E(r)$ are included in the residual
\beq
\delta {\cal E}(R)=
2\pi \int_0^R \der r\,r^2\,\srho(r)\,\spot(r)\left[ \left\lav\frac{\delta\rho}{\srho}\frac{\delta \Phi}{\spot}\right\rav(r)+\frac{\sigma^2(r)-s^2(r)}{\spot(r)}\right]\,.
\label{de}
\eeq
Specifically, the part of it harbouring the difference
$\sigma^2(r)-s^2(r)$ corrects the spherical total energy in the sphere
from the gravitational energy exchanged (it may be positive or
negative) at a distance with the rest of the system, and the other
part with the crossed density--potential fluctuation profile corrects
it for the different potential energy due to the actually
non-spherically symmetric mass distribution of the system.

As virialisation is driven by the energy lost through shell-crossing
(not by the energy exchanged at a distance between shells, which is a
consequence of the former effect and much less important), the ratio
between the final and initial total energies in the sphere is, to
leading order in the deviation from spherical symmetry, equal to the
ratio between the corresponding spherical total energies also
accounting for that energy loss
\beq
\frac{E(R)}{E\p(R\p)}=\frac{{\cal E}(R)}{{\cal E}\p(R\p)}\,.
\label{energies}
\eeq

We stress, once again, that equation (\ref{energies}) is valid to
leading order in the deviation from spherical symmetry. This does not
mean that the total energy in the sphere is approximated by its
spherical counterpart. (By doing this we would lose all the
information on the shape of the system.) It is the whole ratio between
the total energies in the final and initial spheres which is
approximated by the ratio between their spherical counterparts. Thus,
taking into account the relation $E(R)={\cal E}(R)+\delta{\cal E}(R)$,
equation (\ref{energies}) implies
\beq
\frac{{\cal E}(R)}{{\cal E}\p(R)}=\frac{\delta
  {\cal E}(R)}{\delta{\cal E}\p(R)}\,,
\label{cale}
\eeq
to leading order in the deviation from spherical symmetry.

Substituting the expressions for $\delta{\cal E}$ and $\delta{\cal
  E}\p$ in the virialised object and its seed (eq.~[\ref{de}])
into the relation (\ref{cale}) and differentiating it, we are led to
\beq
\frac{\lav\Phi\rav(r)}{{\cal D}(r)}\left[\left\lav\frac{\delta\rho}{\srho}\frac{\delta\Phi}{\spot}\right\rav(r)+\frac{\sigma^2(r)-s^2(r)}{\spot(r)}\right]=\lav\Phi\p\rav(r\p)\left\lav\frac{\delta\rho\p}{\lav \rho\p\rav}\frac{\delta\Phi\p}{\lav \Phi\p\rav}\right\rav(r\p)\,.
\label{1st}
\eeq
where ${\cal D}(R)\equiv {\cal E}(R)/{\cal E}\p(R)$, the so-called
``spherical'' energy dissipation factor, is given by (eqs.~[\ref{vir0}] and [\ref{diff}])
\beq
{\cal D}(R)=\frac{10 R^2}{3GM^2(R)}\int_0^R \der r\left[4\pi\srho(r)GM(r)+\frac{{\cal W}(r)}{2r^2}\right]\,.
\eeq
Then, taking into account equations (\ref{4th}) and (\ref{D1}), both for
the virialised object and its seed, equation (\ref{1st}) establishing
the link between the respective mean crossed density--potential
fluctuation profiles leads to
\beqa
U(r)\!\left\{1\!-\!\frac{3[(1-\ep^2)^2(1-\es^2)^2+(1-\ep^2)^2+(1-\es^2)^2]}
{[(1-\ep^2)(1-\es^2)+(1-\ep^2)+(1-\es^2)]^2}\!-\!S\right\}\!(r)
\!=\!\!\left\{1\!-\!
\frac{3[(1-\ep^2)^2(1-\es^2)^2+(1-\ep^2)^2+(1-\es^2)^2]}
{[(1-\ep^2)(1-\es^2)+(1-\ep^2)+(1-\es^2)]^2}\right\}\!(r\p),
\label{7th}
\eeqa
where
\beq
U(r)\equiv \frac{\spot(r)V\p(r\p)}{\lav\Phi\p\rav(r\p){\cal D}(r)V(r)}~~~~~~~~~~~~~{\rm and}~~~~~~~~~~~~~~~~
S(r)=\frac{5}{2}\frac{\sigma^2(r)-s^2(r)}{\spot(r)}V(r)\,,
\label{S}
\eeq
being
$V(r)=1-\xi(r)\gamma(r)\left\{1-[1+2\kappa]\gamma(r)-\frac{\der\ln
  \gamma}{\der\ln r}\right\}$ and $\gamma(r)$ the logarithmic
derivative of the mean crossed density--potential fluctuation profile.

\begin{figure}
\centerline{\includegraphics[scale=0.43]{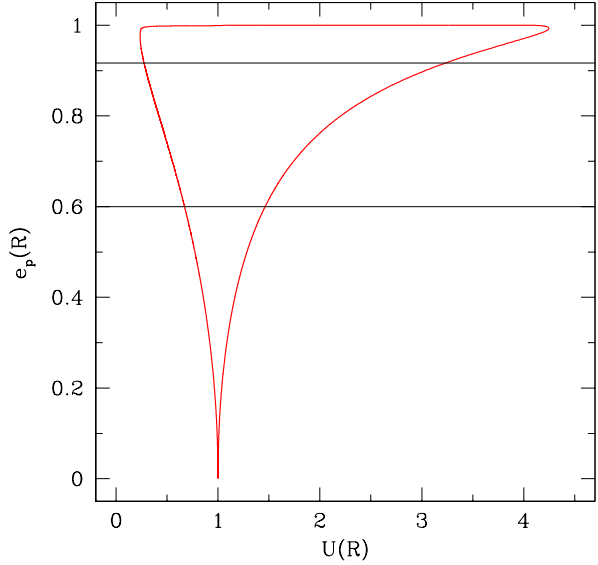}}
\caption{Primary eccentricity space (delimited by the solid red line)
  possible to occupy at small and intermediate radii by current CDM
  haloes with arbitrary mass in the concordance model for varying
  values of the quantify $U(R)$ (see text). The two horizontal black
  lines delimit the values of $\ep$ reported in the literature for
  simulated haloes.}\label{f1}
\end{figure}

Equations (\ref{6th}) and (\ref{7th}) determine the eccentricities of
the virialised object grown by PA from those of its seed, as
wanted. From equations (\ref{7th}) and (\ref{S}), we see that this
relation involves the profile $\sigma^2(r)$, which depends on the
shape of the final system and, hence, on that of its seed. But such a
dependence cannot be determined without previously determining the
anisotropy profile. This will be done in Section
\ref{anisotropy}. Only at small $r$, the term $S(r)$ in equation
(\ref{7th}) becomes negligible\footnote{The $\sigma^2(r)$ profile at
  every radius is of the order of the squared circular velocity,
  $GM(r)/r$, while $|\spot(r)|$ becomes much larger than $GM(r)/r$ at
  small radii.}. This means that, except for the factor $U(r)$,
equation (\ref{7th}) takes there the form of an identity relation,
just as equation (\ref{6th}). Interestingly, the set of algebraic
equations (\ref{7th}) and (\ref{6th}) is solvable only for a very
narrow range of $U(r)$ values around unity. This is shown in Figure
\ref{f1}, where we plot the solution space for varying values of
$U(r)$ inferred by means of the algebraic solving procedure provided
in the Appendix. Consequently, the eccentricities of the virialised
object near the centre are necessarily close to those of the
protoobject at the same location.

\section{KINEMATIC PROFILES}\label{anisotropy}

As mentioned, the kinematics of a non-rotating triaxial virialised
object must depend on its shape and conversely. This relation is
however hard to establish from first principles, even under the
assumption of PA as made here. It can nonetheless be guessed from
comparison with the simplest idealised case of strict spherical
symmetry. See also Section \ref{haloes} for a more fundamental
justification.

In that idealised case, the final velocity dispersion would be purely
radial because the system would expand, collapse and virialise
radially (shell-crossing would take place between spherical
shells). The total energy within spheres is, to leading order in the
dimensionless fluctuation profiles measuring the deviation from
spherical symmetry, the same in spherically symmetric systems as in
triaxial ones (see eq.~[\ref{ener}]). Thus, non-radial infall due to
non-spherical shells (or to any possible instability; see below)
should cause the transfer from radial to tangential kinetic energy
together with a deviation of the gravitational potential from its
spherical average without altering, to leading order, the total energy
of the system. Assuming one-to-one correspondence between the two
effects, we are led to the conclusion that, to leading order, the
fractional velocity variance transferred from the radial to the
tangential direction (equal to half the fractional 1-D tangential
velocity variance generated) should be equal to half the typical
fractional deviation of the potential from its spherical average,
\beq 
\frac{\sigma^2\tang(r)}{\sigma^2(r)}=\left\lav\left(\frac{\delta\Phi}{\spot}\right)^2\right\rav^{\!1/2}\!\!\!(r)\,.
\label{00th}
\eeq
Equation (\ref{00th}) might seem to imply that it is impossible to
have tangential velocities in spherically symmetric virialised objects
(with null $\delta\Phi$), which would be manifestly wrong. What it
actually implies is this very conclusion but {\it for objects formed
  by PA}\footnote{In Section \ref{mm}, we will see that it also holds
  for virialised objects having undergone major mergers, although, in
  this particular case, such an implication is quite
  obvious.}. Indeed, as discussed in SVMS, such objects are always
ellipsoidal (they have non-null $\delta\Phi$). Even if the seeds are
spherically symmetric, the gravitational effect of the surrounding
matter (non-null shear tensor) or radial orbit instability, in the
artificial case of isolated seeds, automatically leads to non-radial
collapse and the formation of triaxial virialised objects (see
Sec.~\ref{haloes} for more details).

Given the relation (\ref{00th}), the anisotropy profile,
\beq 
\beta(r)=
1-\frac{\sigma\tang^2(r)}{\sigma\rad^2(r)}=\frac{1}{2}\left[3-
\frac{\sigma^2(r)}{\sigma^2\rad(r)}\right]=1-\frac{\frac{\sigma\tang^2(r)}{\sigma^2(r)}}{1-2\frac{\sigma\tang^2(r)}{\sigma^2(r)}}\,,
\label{beta}
\eeq
is a function of the rms potential fluctuation profile for the
virialised object. As shown in Section \ref{eccentricity}, such a
fluctuation profile is related to its counterpart in the
seed. However, such a relation involves the velocity dispersion
profile itself (eq.~[\ref{7th}]). Thus, $\sigma^2(r)$ must be
determined before the rms potential fluctuation profile.

To do this we will make use of the (exact) generalised Jeans equation
for steady non-spherical virialised objects (see SVMS), which, taking
into account equation (\ref{beta}), can be written as
\beqa
\frac{\der [(3-2\beta)^{-1}\srho\, \sigma^2]}{\der r}\!+\!
\frac{2\beta(r)}{3-2\beta(r)}\frac{\srho(r)\sigma^2(r)}{r}
\!+\!\srho(r)\,\frac{GM(r)}{r^2}
=-\srho(r)\spot(r)
\!\left[\frac{2}{r}\left\lav
\frac{\delta\rho}{\srho}\frac{\delta\Phi}{\spot}\right\rav(r)
\!+\!\left\lav
\frac{\partial}{\partial r}\!\left(\frac{\delta\rho}{\srho}\right)\!\frac{\delta\Phi}{\spot}\right\rav(r)\right]\!.
\label{exJeq2}
\eeqa
The member on the right of equation (\ref{exJeq2}) is of second order
in the dimensionless fluctuation profiles. Like $\beta(r)$, these
profiles are functions of their counterparts in the seed and of
$\sigma(r)$. Consequently, equation (\ref{exJeq2}) is a differential
equation for $\sigma(r)$, which can be solved for the usual boundary
condition of null dispersion at infinity\footnote{The inside-out
  growth of object formed by PA guarantees that the solution out to
  any radius is kept unaltered as the radius of the object increases,
  so we have indeed the right to endorse the boundary condition at
  infinity.}. Moreover, the member on the right of equation
(\ref{exJeq2}) is of second order in the deviations from spherical
symmetry, while all remaining terms are of lower order; the anisotropy
profile itself is correct only to first order. Therefore, it can be
neglected in equation (\ref{exJeq2}), which notably simplifies the
solution of this equation. Once $\sigma(r)$ has been determined, the
eccentricities of the virialised object can be calculated from
equations (\ref{6th}) and (\ref{7th}) and $\beta(r)$ from equation
(\ref{beta}).

\section{APPLICATION TO CDM HALOES}\label{haloes}

In SVMS, the spherically averaged density profile for CDM haloes was
shown to arise from the spherically averaged density profile of halo
seeds. As this latter profile is determined by the power-spectrum of
density perturbations, it was possible to infer the typical halo
density profile for CDM haloes directly from the power-spectrum of
density perturbations. The extension of the model presented in the
present paper relates the shape and kinematics of CDM haloes to the
shape of their seeds, which is also determined, of course, by the
power-spectrum of density perturbations. Thus, it should also be
possible to infer the typical shape and kinematics of haloes directly
from the power-spectrum of density perturbations.

To do this we need to know the eccentricity profiles for protohaloes,
while what is only known is the typical eccentricities at peaks
(i.e. at $r\p=0$) in the initial density field {\it filtered} by a
Gaussian window, calculated by BBKS. Nonetheless, the former can be
accurately inferred from the latter, following the procedure developed
in SVMS for the spherically averaged density profile.

As pointed out in SVMS, the fact that haloes grown by PA {\it develop
  from the inside out} implies that the density profile for each halo
ancestor in the continuous series leading to the final object exactly
matches that for the halo within one triaxial isodensity contour. All
these halo ancestors evolve from peaks in the initial density field
filtered at the ancestor scale. Thus, the typical spherically averaged
density contrast profile for the protohalo, $\lav \delta\p\rav (r\p)$,
convolved by a Gaussian window with any radius $\R$ must be equal to
the typical density contrast of a peak, $\delta\pk(\R)$, in the
density field filtered at scale $\R$,
\beq 
\delta\pk(\R)=
\left(\frac{2}{\pi}\right)^{1/2}\R^{-3}\,\int_0^\infty \der
r\p\,r\p^2\,\lav \delta\p\rav
(r\p)\exp\left(-\frac{r^2\p}{2\R^2}\right).
\label{Fred}
\eeq 
The typical trajectory $\delta\pk(\R)$ followed by peaks evolving by
PA into the series of ancestors leading to a typical halo with $M$ at
$t$ is the solution of the differential equation \citep{MSS96},
\beq
\frac{\der \delta\pk}{\der \R}= -x_{{\rm e}}(\R,\delta\pk)\,\sigma_2(\R)\,\R\,,
\label{dmd}
\eeq
for the appropriate boundary condition (through the $\delta\pk(t)$ and
$\R(M)$ relations given by eqs.~[40] and [41] of SVMS). In equation
(\ref{dmd}), the inverse of $x_{{\rm e}}(\R,\delta\pk)$ is the mean
inverse curvature $x$ (i.e. minus the Laplacian over the second order
spectral moment) of peaks with $\delta\pk$ at $\R$. The differential
equation (\ref{dmd}) can be solved, which then allows one to inverse
the Fredholm integral equation (\ref{Fred}) and obtain the wanted
density contrast profile for the typical protohalo (see SVMS for
details). This procedure was followed in SVMS to infer the unconvolved
spherically averaged density profile for the seed of a typical halo.

In the present case, we must take into account that the squared
semiaxes of each peak over the peak trajectory leading to the typical
halo with $M$ at $t$ coincide (except for an arbitrary scaling) with
minus the second order spatial derivatives $\lambda^2\jj$ of the
filtered density contrast at the peak, in Cartesian coordinates $x\jj$
with $x_1$ aligned along the major semiaxis. They are therefore
related to the density contrast profile, $\delta\p({\bf r})$, for the
protohalo through
\beq 
\lambda^2\jj(\R,\delta\pk)\,\sigma_2(\R) =\frac{-1
}{(2\pi)^{3/2}\R^3}\left\{\frac{\partial^2}{\partial x\jj^2}\int \der
  {\bf r}\p\, \delta\p({\bf r}\p)\,\exp\left[-\frac{1}{2}\left(\frac{{\bf r}\p-{\bf
        x}}{\R}\right)^2\right]\right\}_{{\bf x}=0} \!.
\label{lambda}
\eeq
Writing the density contrast profile in the halo in terms of its
spherical average (this relation is similar to that between the
respective total densities, given by eqs.~[\ref{ellips}] and
[\ref{norm}]), performing the second order spatial derivatives of the
Gaussian window inside the integral on the right of equation
(\ref{lambda}) and taking ${\bf x}=0$, changing to spherical
coordinates, integrating over $\theta$ and $\phi$ and averaging over
the peak curvature, we arrive at
\beqa
\left(\frac{\pi}{2}\right)^{1/2}\,\R^3\left[\delta\pk-\frac{A\jpk^2\xbra}{3}(\R,\delta\pk)\,\sigma_2(\R)\,{\R^2}\right]-\frac{1}{5{\R^2}}\int_0^\infty \der
r\p\,r\p^4\,\lav\delta\p\rav(r\p)\,\exp\left(-\frac{r\p^2}{2\R^2}\right)\nonumber\\
=\frac{2}{5\R^2}\int_0^\infty \der
r\p\,r\p^4\,\lav\delta\p\rav(r\p\,)\frac{\,a\p^2(r\p)}{a\jp^2(r\p)G\p(r\p)}\,\exp\left(-\frac{r\p^2}{2\R^2}\right),~~~~~~~~~~~~~~~~~~~~~~~~~~~~
\label{inv}
\eeqa
where $\xbra(\R,\delta\pk)$ is the mean curvature of peaks with
$\delta\pk$ and $\R$ and $A\jpk(\R,\delta\pk)$ (j $=1\div 3$) are its
associated dimensionless semiaxes $\lambda\jj$ scaled to the square
root of the Laplacian, so that they satisfy
$A\opk^2+A\tpk^2+A\zpk^2=1$. These semiaxes depend on $\delta\pk$ and
$\R$ through $\xbra$ in a well-known form calculated by BBKS. Thus, we
can calculate them as well as the curvature $\xbra$ over the peak
trajectory $\delta\pk(\R)$ leading to the halo with $M$ at $t$ and
invert equation (\ref{inv}) by means of the same procedure as used for
equation (\ref{Fred}). In addition, the integral on the left of
equation (\ref{inv}) can also be calculated from the known spherically
averaged protohalo density contrast profile. Thus, taking the ratios
between the solutions for different j obtained from inversion of
equation (\ref{inv}), we can infer the eccentricity profiles for the
seeds of typical haloes grown by PA.

Once the eccentricity profiles, $(\ep)\p(r\p)$ and $(\es)\p(r\p)$, are
known, the typical halo shape and kinematics can be inferred following
the steps described in Sections \ref{eccentricity} and
\ref{anisotropy}: 1) infer (through eqs.~[\ref{6th}] and [\ref{7th}])
the eccentricity profiles $\ep(r)$ and $\es(r)$ of the halo as a
function of $\sigma(r)$; 2) determine from them the mean squared
density fluctuation profile for the halo (eq.~[\ref{4th}]) as a
function of $\sigma(r)$; 3) infer the corresponding mean squared
potential fluctuation profile (eqs~[\ref{D1}] and [\ref{D3}]) as a
function of $\sigma(r)$; 4) solve the generalise Jeans equation
(\ref{exJeq2}) for the velocity dispersion profile $\sigma(r)$; and 5)
determine (from eqs.~[\ref{6th}] and [\ref{7th}]) the typical
eccentricity profiles, $\ep(r)$ and $\es(r)$, now using the explicit
values of $\sigma(r)$, and the anisotropy profile $\beta(r)$
(eq.~[\ref{00th}] after deriving the mean squared potential
fluctuation profile).

Unfortunately, the accurate theoretical eccentricities so inferred
could not be compared with the results of numerical simulations
because there is so far no such typical profiles drawn from
simulations. We just know some main trends of the typical shape of
haloes. Thus, it is not worth at this stage to carry out such a
complex derivation. In fact, the purpose of the present paper is not to infer
accurate eccentricity and kinematic profiles for CDM haloes directly
from the power-spectrum of the concordance model but rather to verify
the validity of the model and try to understand the origin of the
universal trends shown by such halo properties in numerical
simulations. And these two objectives are much better obtained by
means of the following {\it simpler and comprehensive} approximate
procedure.

Peaks are triaxial with a rather prolate shape (BBKS). As mentioned,
their eccentricities depend on $\delta\pk$ and $\R$ through their
curvature $x$ in such a way that, the larger $x$, the more spherical
is the peak. The quantity $\xbra$ increases progressively with
increasing $\R$ over typical peak trajectories, meaning that peaks
become increasingly spherical over those tracks, very slowly first and
much more rapidly at the end. Thus, the same trend must be found in
protohaloes for increasing $r\p$ as well as in typical haloes for
increasing $r$. (It is true that the relation between the
eccentricities in the seed and the halo also depends on $\sigma(r)$,
but this dependence can be neglected at small radii, while at large
radii the decrease with increasing $r\p$ of protohalo eccentricities
is so marked that the dependence on $\sigma(r)$ cannot reverse it.)
Therefore, according to the present model, haloes should be rather
prolate and approximately homologous at small and intermediate radii
and tend to become more spherical near the halo edge. This behaviour
is fully consistent with the results of numerical simulations
(e.g. \citealt{JS02,bs05,All06,St09}).

Moreover, we can have an idea on the overall values of halo
eccentricities because, as mentioned, they are expected to be rather
uniform except near the edge. The typical ratio between the major and
minor axes (the major and intermediate axes) in peaks with low and
moderate average curvature is $\sim 1.7$ ($\sim 1.3$) (BBKS), implying
typical values of the eccentricities $\ep$ ($\es$) of $\sim 0.81$
($\sim 0.64$). Such typical values, independent of the filtering scale
(they only show a moderate dependence on the curvature) were derived
by BBKS for peaks in the old standard CDM model (i.e. in the
Einstein-de Sitter cosmology), so they are valid for any CDM cosmology
as all cosmologies converge for high enough redshifts as those
corresponding to primordial peaks to the Einstein-de Sitter model. As
$U(r)$ is close to one with a small bias towards two at small $r$,
the expected values of $\ep$ ($\es$) in haloes at small radii should
be close to, perhaps a little larger than, those of peaks, say, $\sim
0.9$ ($\sim 0.8$). This is also consistent with the values found in
numerical simulations
\citep{Fea88,Bull02,JS02,Sp04,KE05,bs05,All06,Ha07,Be07,St09}.

\begin{figure}
\vskip -0.7 cm
\centerline{\includegraphics[scale=1.]{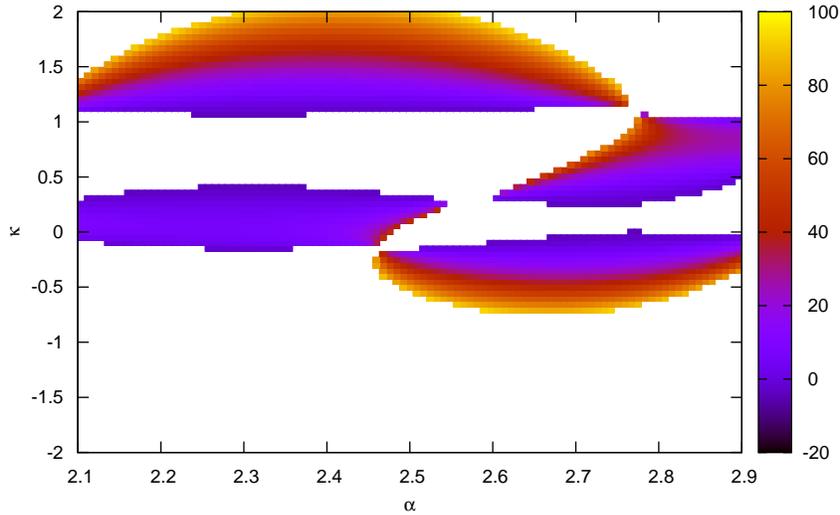}}
\caption{Same as Figure 1 for asymptotic power-law regimes with the
  range of $\kappa$ and $\alpha$ indexes leading to only moderate
  values (not much greater than one) of $Q/P$ as found in
  numerical simulations.}\label{f3}
\end{figure}

At large $r$, the dependence on $\sigma^2(r)$ of the relation between
the eccentricities in the halo and the seed becomes important.  Yet,
we can still determine the expected outer asymptotic behaviour of the
eccentricities in the halo, taking only into account that, as
mentioned, the object becomes more spherical at very large $r$. In
that asymptotic regime, the power indexes of the mean squared density,
squared potential and crossed density--potential fluctuation profiles
coincide (see Sec.~\ref{eccentricity}) and the more or less marked
departure of isodensity and isopotential contours from spheres, given
by the ratio $Q/P$, depends on the index $\kappa$ and the asymptotic
logarithmic slope $-\alpha$ of the density profile
(eq.~[\ref{QP}]). For values of $\alpha$ approaching 3 as near the
outer asymptotic regime of virialised haloes, the only possible
negative values of $\kappa$ leading to outward-decreasing
eccentricities with isopotential contours moderately more spherical
than the isodensity contours ($Q/P$ not much greater than one) as
found in the literature \citep{Sp04,KE05,Ha07} are in the range
$-0.2\la \kappa< 0$ (see the regions in dark blue, or quite strong
grey if black and white, in Fig.~\ref{f3}). Thus, the outer
asymptotic logarithmic slope, $\kappa$, is also severely constrained
in this approximate treatment.

\begin{figure}
\centerline{\includegraphics[scale=0.43]{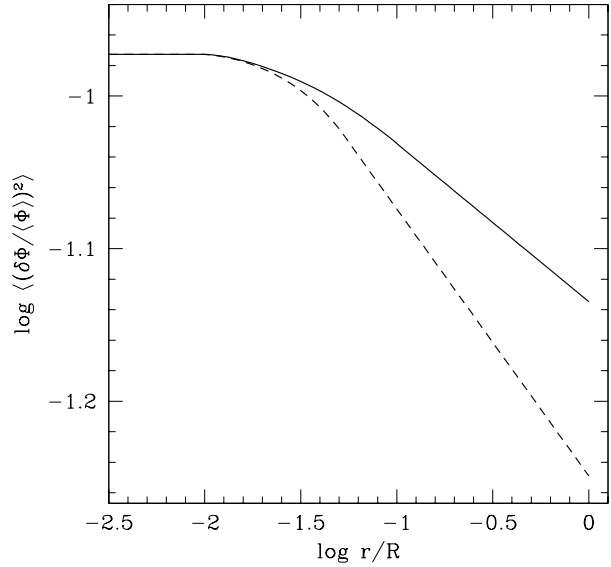}}
\caption{Plausible typical mean squared potential fluctuation profiles
  according to our model. They connect in a smooth way an inner
  asymptotic uniform value consistent with those found in Figure
  \ref{f1} and outer asymptotic logarithmic slope $\kappa$ equal to
  $-0.1$ (solid line) and $-0.175$ (dashed line) within the allowed range
  according to Figure \ref{f3}.}
\label{fIII2}
\end{figure}

And what about the typical halo kinematics? The typical mean squared
potential fluctuation profile for CDM haloes in log-log should be
approximately given by the natural spline between a uniform value of
$\sim 0.11$, as implied by the above mentioned typical eccentricities,
below about one hundredth the virial radius and a straight line with
logarithmic slope $\kappa$ beyond about one tenth the virial
radius. We will consider two values of $\kappa$, $-0.175$ and $-0.1$,
in order to better sample the allowed range $-0.2\la \kappa< 0$ and to
see how robust the theoretical typical halo kinematics are against
variations in the individual shapes of haloes. (Different triaxial
shapes must give rise to different mean squared potential fluctuation
profiles.) The two squared potential fluctuation profiles so built are
shown in Figure \ref{fIII2}.

\begin{figure}
\vskip -70pt
\centerline{\includegraphics[scale=.60]{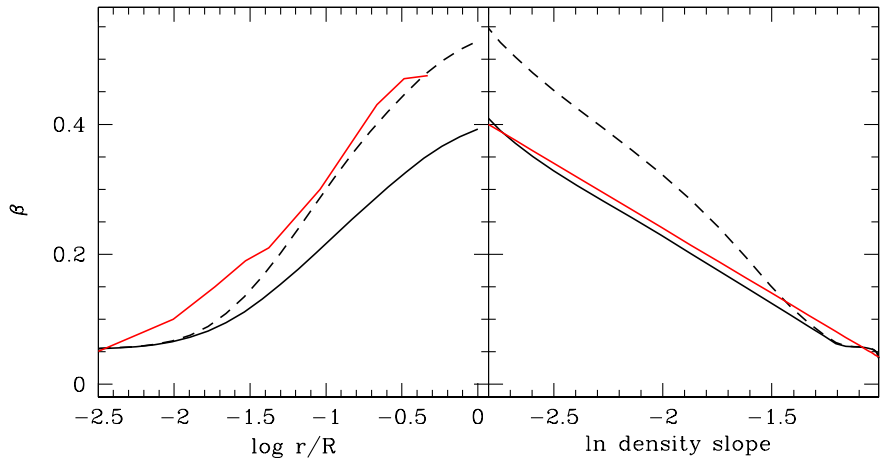}}
\caption{Theoretical anisotropy profile derived from the rms potential
  fluctuation profiles shown in Figure \ref{fIII2} (same lines). {\it
    Left panel:} as a function of radius to facilitate the comparison
  with the curve found for the Milky Way mass halo (red line)
  simulated by \citeauthor{Navea10} (\citeyear{Navea10}; see their Fig.~10). 
  {\it Right panel:} as a function of the logarithmic density slope to
  facilitate the comparison with the analytical formula by \citet{HS06} 
  fitting the anisotropy profiles for simulated haloes
  (red line).}
\label{fIII5}
\end{figure}

Once the form of the squared potential fluctuation profile has been
fixed, we can calculate the ratio $\sigma\tang^2(r)/\sigma^2(r)$
(eq.~[\ref{00th}]) and, from it, the anisotropy profile
(eq.~[\ref{beta}]). Figure \ref{fIII5} shows the velocity anisotropy
profiles inferred from the two different squared potential fluctuation
profiles. As can be seen, they are both in good agreement with the
results of numerical simulations. Specifically, for $\kappa=-0.1$, we
find an anisotropy profile that closely follows the typical
`universal' law proposed by \citet{HS06} (the red line in the right
panel). While, for $\kappa=-0.175$, we find it closer to the
anisotropy profile found by \citet{Navea10} in a realisation of a
Milky Way mass halo (the red line in the left panel). The only
noticeable disagreement in this latter case is that the theoretical
profiles show no cutoff near the halo edge as found by
\citeauthor{Navea10} But this is a very unusual feature (it does not
follow the `universal' trend) and likely reflects an incomplete
relaxation of this particular simulated halo at those large radii.

\begin{figure}
\centerline{\includegraphics[scale=.43]{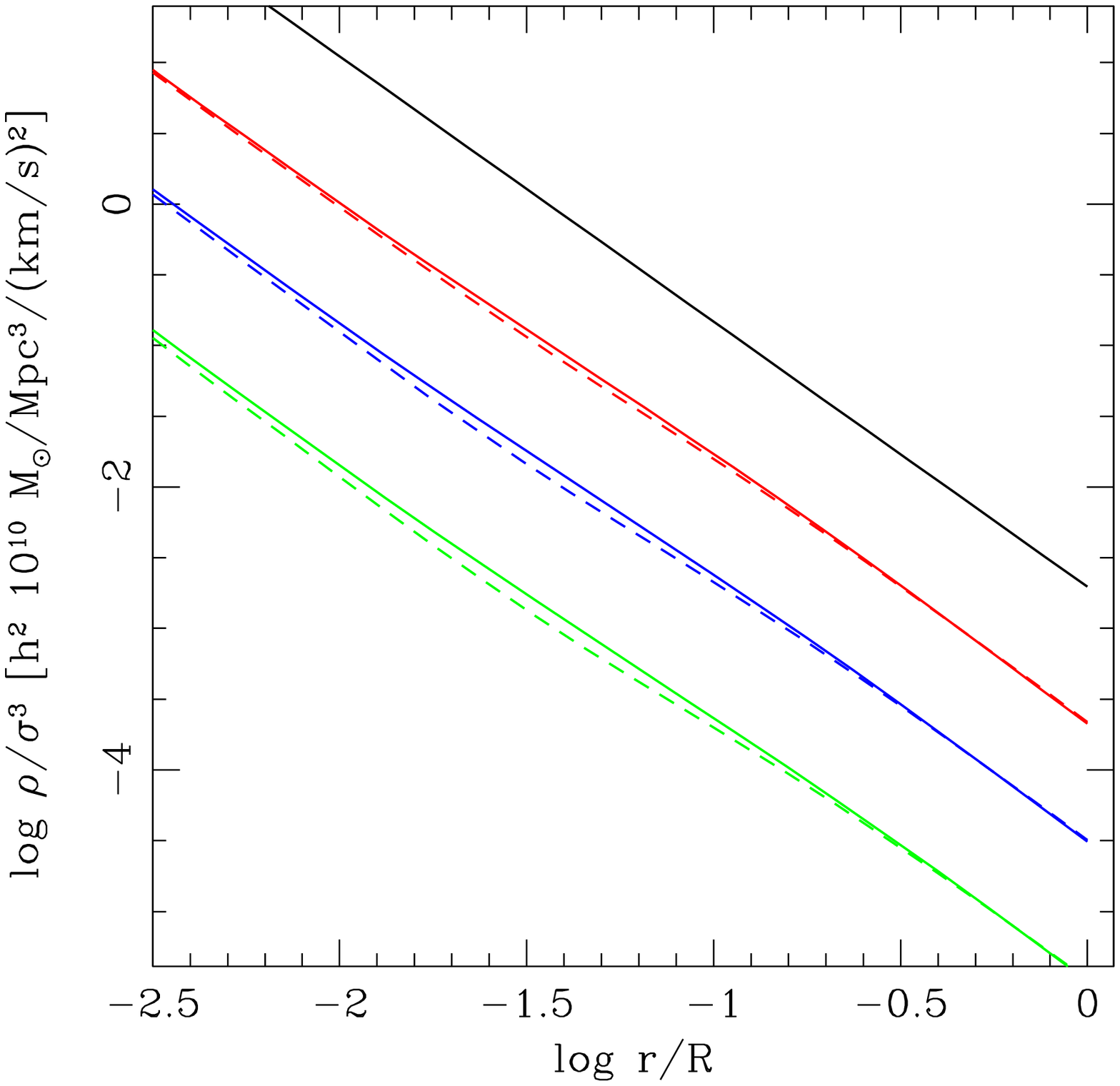}}
\caption{Theoretical pseudo phase-space density profile resulting from
  the two anisotropy profiles shown in Figure \ref{fIII5} (same
  lines), compared to a straight line with slope $-1.875$ and
  arbitrary zero-point as found in simulations (black line). For
  comparison we also plot the theoretical profiles obtained for haloes with
  masses equal to $10^{13}$ \modot (blue lines) and $10^{14}$ \modot
  (green lines).}
\label{f6}
\end{figure}

Using the previous theoretical velocity anisotropy profiles, we can
solve the generalised Jeans equation (\ref{exJeq2}) to find the
corresponding velocity dispersion profiles and, using the typical
spherically averaged halo density profile derived in SVMS, the
corresponding pseudo phase-space density profiles,
$\srho(r)/\sigma^3(r)$. In Figure \ref{f6}, we show the results
obtained from the two mean squared potential fluctuation profiles
above. In both cases, the theoretical pseudo phase-space density
profile is very nearly a power-law with a logarithmic slope of
$-1.875$, in full agreement with the results of numerical
simulations. The two solutions essentially overlap; there is just a
slight undulation at intermediate radii in the solution obtained from
$\kappa=-0.175$, which indicates that the spline carried on between
the inner and outer asymptotes of the squared potential fluctuation
profile is somewhat deficient in this case. Nonetheless, the fact that
so distinct mean squared potential fluctuation profiles as adopted
here give so similar solutions demonstrates that the pseudo
phase-space density profile is very robust against variations in halo
shape (see Sec.~\ref{mm} for the possible origin of this
  behaviour). Note instead the clear dependence of the zero-point of
the pseudo phase-space density profile on the mass of the halo
predicted by the model. This cannot be compared with simulations as
this particular dependence has not yet been analysed. It would be
worth trying to confirm this prediction of the model.

According to our results, the velocity anisotropy profile should show
an important dependence on the shape of the system, which is
consistent with the rather large scatter shown by this profile in
numerical simulations. The velocity dispersion profile follows from
the Jeans equation (\ref{exJeq2}) for such an anisotropy profile and
the density profile $\srho(r)$. As the latter profile is independent
of the shape of the object, it cannot balance in the Jeans equation
the dependence of $\beta(r)$ on that property. Consequently,
$\sigma(r)$ must also depend on the shape of the object. Yet, the
pseudo phase-space density profile turns out to be very robust against
variations in that property; there are just very slight undulations
depending on it. Consequently, $\sigma(r)$ must also be similarly
robust. This is once again in agreement with the small scatter shown
by this profile in simulated haloes.

To end up we want to remark that we have made so far no reference to
the CDM cosmology used; all the preceding (approximate) results hold,
regardless of the cosmology provided only haloes grow hierarchically
from the bottom-up (see SVMS). This is once again in agreement with
the results of numerical simulations. Of course, the finer details of
the predicted profiles will depend on the specific power-spectrum
used, but this will not modify the main properties discussed here.

\section{DISCUSSION}\label{mm}

As just shown, the present model recovers the right form of the
anisotropy and pseudo phase-space density profiles of simulated
haloes. The origin of the former profile is clear: it arises from the
tight relation (\ref{00th}) between the tangential velocity dispersion
and the mean potential fluctuation profile and the typical form
(essentially the well-constrained inner and outer asymptotic
behaviours) of this latter profile. But the origin of the
power-law-like pseudo phase-space density profile is less obvious. On
the other hand, we have assumed so far that haloes grow by PA while
they grow both through PA and major mergers. In this Section, we
address these two important points.

\subsection{Origin of the Pseudo Phase-Space Density Profile}

The form of the pseudo phase-space density profile of simulated
haloes, recovered by the present model, coincides with that predicted
by Bertschinger's (1985) model for collisionless spherically
symmetric, self-similar, systems evolving by PA. This implies that the
ultimate reason for such a coincidence cannot be neither the shape
(spherically symmetric vs. triaxial) of the system nor the form
(power-law vs. non-power-law) of the density profile. The only more
fundamental aspect in the formation of virialised objects grown by PA
directly or indirectly included in both models is the way the
coarse-grained phase space density increases via phase mixing (entropy
generation) when the system loses energy via shell-crossing and
contracts.

Bertschinger's model is dynamical and can properly follow the effects
of shell-crossing, while the SVMS model is steady and focuses on the
equilibrium state resulting from that process. However, the inside-out
growth condition the SVMS model relies on directly follows from the
lack of apocentre-crossing during virialisation, a condition that is
also fulfilled by Bertschinger's model. Thus, despite the distinct
approach, the effects of shell-crossing are similarly accounted for in
the two models.

Certainly, the power-law form of the pseudo phase-space density
profile predicted by Bertschinger's model is tightly related to the
self-similarity assumption, while the present model makes no such an
assumption. However, PA is always close to self-similar. It is driven
by gravitation, a scale-free force, and the initial conditions are
also very approximately scale-free: the initial mass distribution is
essentially uniform (with essentially the critical density) and the
velocity field corresponds to the scale-free unperturbed Hubble-flow
form. Thus, the solution of the present model at small $z$ should not
be very far from self-similar, except for the effects of the
increasing departure from the Einstein-de Sitter universe owing to the
specific density profile of the seed. In this sense, the predicted
profiles should not be far from power-laws, particularly the pseudo
phase-space density profile that appears to be so little sensitive to
the details of the seed. Another fundamental assumption of
Bertschinger's model is the radial infall of the system, while the
SVMS model takes into account that the collapse and virialisation is
actually non-radial. However, if the tangential velocities develop at
the expense of the initial radial velocities without altering the
total velocity dispersion, as assumed in equation (\ref{00th}), the
velocity dispersion that emerges as a consequence of shell-crossing
should be the same in spherically symmetric as in triaxial systems.

The situation is therefore as follows. The power-law form, with index
equal to -1.875, of the pseudo phase-space density profile found by
Bertschinger is the direct consequence of the shell-crossing produced
in self-similar radial PA. Thus, the conditions that: i) {\it PA is
  always quite close to self-similar} and ii) {\it the velocity
  dispersion generated by shell-crossing is not modified}, according
to the condition (\ref{00th}), by the appearance of tangential
velocities in non-radial PA, should guarantee that the evolution of
the coarse-grained phase-space density found in Bertschinger's model
is kept essentially unaltered in the general (non-striclty
self-similar and non-radial) case of PA. This would explain why the
present model recovers the pseudo phase-space density profile \`a la
Bertschinger in agreement with the results of numerical
simulations. Note that such a pseudo phase-space density profile does
not necessarily implies that the density profile must also be
roughly proportional to $r^{-2.25}$ as found by Bertschinger. This
would be the case if PA were strictly self-similar, so to end up with
a power-law density profile, and, more importantly, radial. As stated
by \citet{Ber85} when comenting on his density profile $\rho\propto
r^{-2.25}$: ``Self-similar relaxation is not complete [as it proceeds
along one direction only] and thus heuristically should not lead to a
density profile as flat as $\rho \propto r^{-2}$ [as predicted by
\citet{LB67} in the case of complete 3D relaxation]. It is uncertain
to what extent the addition of angular momentum (i.e. non-radial
orbits) will change this result,...''. In other words, if PA is
non-radial, than the power index of the density profile is expected to
change, in agreement with the results of SVMS, despite the fact that
the pseudo-phase space density profile, driven by the evolution of the
phase-space density during virialisation in the absence of
apocentre-crossing, should be kept roughly unaltered as a consequence
of condition (\ref{00th}).

As mentioned, a key point in the present model, arising from the
particular way shell-crossing proceeds in accreting virialised haloes,
is their inside-out growth. Bertschinger's model is instead based on
the self-similar evolution of such systems and does not care about the
implications this has on the growth of the steady object. But, if our
explanation of the power-law-like pseudo phase-space density profile
is correct, the two models should essentially coincide (except for the
symmetry and the strict self-similarity condition), meaning that
Bertschinger's solution should also approximately satisfy the
inside-out growth condition. This is indeed the case. As stated by
Bertschinger, as a consequence of shell-crossing, particle orbits
rapidly become approximately periodic, that is, the mass inside them
stabilises and the inner system does not change anymore. In other
words, the system develops a steady core and grows from the inside-out
as new shells virialise. Of course, such a behaviour can only be {\it
  approximately} satisfied in Bertschinger's model. As a consequence of 
self-similarity, steadiness can only be strictly achieved in the {\it
  limit of infinite time} or {\it vanishing radii}; particle orbits can
never become exactly periodic, they can only tend to become so (see
Betschinger's Fig.~9 at the base of his comment regarding the
inside-out growth). To obtain a fully steady solution, the
self-similarity assumption must be replaced by the inside-out growth
condition as in the present model.

\subsection{Major Mergers}

As has long been known, there is in radial PA a one-to-one
correspondence between the density profiles of virialised objects and
their seeds \citep{Ber85,DPea00}. The reason for this is that in
radial PA, there is no phase-mixing of particles turning around at the
same time \citep{Ber85}) or, equivalently, there is no
apocentre-crossing of particle orbits during virialisation. This
result was extended in SVMS to non-radial PA.

An important consequence of this result was that the spherically
averaged density profile $\srho(r)$ for a virialised halo does not
allow one to tell whether or not it has suffered major
mergers\footnote{We are of course presuming that the halo has
  completed relaxation after its last major merger.}. The reason for
this is that, given a virialised halo with any arbitrary aggregation
history, we can always think about one peak with appropriate density
profile (or equivalently, the spherical energy distribution ${\cal
  E}\p(M)$), leading by PA to a virialised object with identical
spherically averaged density profile. Such a putative accreting seed
of the halo really exists; it is the peak tracing the current
evolution by smooth accretion of the halo according to the peak
formalism. Thus, it is a normal peak contributing to the peak number
density just as the halo associated with it is a normal halo
contributing to the halo mass function. Therefore, the halo is
indistinguishable from one evolved by PA, regardless of its real
aggregation history.

But what about the shape and kinematics of haloes? Do they allow one
to tell between objects formed by PA and having suffered major
mergers? The answer is always the same: major mergers go unnoticed in
these properties as well. The proof is similar to that carried out in
SVMS for the density profile. Using the one-to-one relations inferred
in Section \ref{eccentricity} thanks to the results of SMVS, given a
halo with any arbitrary aggregation history, one can always think
about one peak not only with the appropriate density profile, {\it
  independent of the ellipsoidal shape of the peak}, leading by PA to
the spherically averaged profile of the halo (SVMS), but also with a
shape also leading to its shape and kinematics. The shape of this
putative accreting seed, coincides, by continuity, with that of the
peak tracing the halo in the peak formalism (see SVMS) and is thus a
normal peak contributing to the peak number density. As a consequence,
the halo is indistinguishable from the one that would grow by PA from
that seed not only regarding the spherically averaged density profile,
but also the shape and kinematics. This result can be summarised by
saying that major mergers go unnoticed in virialised objects regarding
all those properties.

As discussed in SVMS, the previous conclusion reflects the fact that
virialisation is a real relaxation process yielding the memory loss of
the halo past history. This obviously affects not only the inner
structure but also the shape and kinematics of virialised
objects. Certainly, the one-to-one correspondence between virialised
haloes and their seeds in PA implies that there is no full memory loss
in PA due to virialisation, in agreement with the results of
simulations (e.g. \citealt{VCea11}). However, the fact that it is
impossible to reconstruct the seed of any simulated virialised halo by
running the simulation backwards is evidence of the existence of a
time arrow indicating the direction of virialisation. What makes both
aspects consistent is the fact that, as just mentioned, the properties
of virialised haloes do not allow one to tell whether they have
suffered major mergers or they have evolved by PA. Therefore, we
cannot unambiguously determine their initial conditions. In other
words, we can reconstruct their putative accreting seeds, but these
may or may not be their real (possibly multinode) seeds.

An important corollary of this result is that the present model for
the shape and kinematics of virialised objects formed by PA also holds
for objects having suffered major mergers. One must simply infer the
properties of the virialised object from those of its putative
accreting seed, that is the peak tracing its current evolution by
smooth accretion.

\section{SUMMARY AND CONCLUSIONS}\label{summ}

The model developed in SVMS for the inner structure (spherically
averaged density profile) of virialised haloes has been extended to
deal with their triaxial shape (eccentricity profiles) and kinematics
(velocity dispersion and anisotropy profiles). To do this we have
considered the simple scenario of haloes evolving by pure
accretion. Under this assumption we have derived the shape and
kinematics of the final objects from the shape of their seeds. The
reason why dealing with the case of pure accretion is sufficient is
that all virialised halos, even those having undergone major mergers,
can be seen to arise in this way from the peak which, according to the
peak formalism, traces their current evolution. The fact that in pure
accretion the final objects keeps the memory of the initial conditions
is a consequence of virialisation being achieved, in this case,
through shell-crossing with no crossing of particle apocentres. This
is not contradictory with the idea that virialisation is a real
relaxation process. On the contrary, the fact that major mergers do
not leave any particular imprint in the shape and kinematics of
virialised objects implies that one cannot be sure about the real
aggregation history or, equivalently, about the real seed of any given
virialised object.

Applied to virialised haloes in hierarchical cosmologies, the model
allows one to make the link between the typical (mean) shape and
kinematics of these objects and the power-spectrum of random Gaussian
density perturbations. Following a simpler comprehensive approximate
approach, we have shown that the overall shape and kinematics
predicted by the model are consistent with the results of simulations
of CDM cosmologies. In particular, the theoretical anisotropy profile
has the same universal form as empirically found, with a substantial
scatter due to the variety of peak (halo) shapes. Likewise, the
theoretical pseudo phase-space density profile has very approximately
a power-law form with the logarithmic slope of $-1.875$ shown by
simulated haloes. This profile is much more robust, that is, it does
not essentially depend on the peak (halo) shape.  

The present model suggests that the origin of the power-law-like form
of the pseudo phase-space density profile of virialised haloes formed
by pure accretion is due to the particular evolution of the
coarse-grained phase-space density during virialisation. This takes
place through the phase mixing produced by shell-crossing in one
direction with no crossing of particle apocentres. The fact that PA is
always very nearly self-similar and that the velocity dispersion is
conserved where a tangential velocity component develops in non-radial
PA would explain why the solution found by \citet{Ber85} for
self-similar, radial PA, with the same kind of phase mixing, is very
nearly preserved.

\vspace{0.75cm} \par\noindent
{\bf ACKNOWLEDGEMENTS} \par
 
This work was supported by the Spanish DGES, AYA2006-15492-C03-03 and
AYA2009-12792-C03-01, and the Catalan DIUE, 2009SGR00217. One of us,
SS, was beneficiary of a grant from the Institut d'Estudis Espacials
de Catalunya.

\appendix

\section{SOLUTION METHOD FOR THE ECCENTRICITIES OF THE VIRIALISED OBJECT FROM THOSE OF THE PROTOHALO}\label{A}

To solve equations (\ref{6th}) and (\ref{7th}), this latter without
the term in $\sigma(r)-s^2(r)$, or, equivalently (see
eq.~[\ref{4th}]), equations (\ref{6th}) and
\beq
-\frac{5}{2}\,U(r)\left\lav \left(\frac{\delta\rho}{\srho}\right)^2\right\rav(r)
= \left\{1-
\frac{3[(1-\ep^2)^2(1-\es^2)^2+(1-\ep^2)^2+(1-\es^2)^2]}
{[(1-\ep^2)(1-\es^2)+(1-\ep^2)+(1-\es^2)]^2}\right\}(r\p)\,,
\label{8th}
\eeq
it is convenient to change the variables $\ep$ and $\es$ into the
quantities $r\e/r$ and $r\e/a$. Given the definition of the
eccentricities (\ref{primary}), the relation between the old and new
variables is
\beq
\left(\frac{r\e}{a}\right)^6=(1-\ep^2)(1-\es^2)~~~~~~~~~~~~~~~{\rm and}~~~~~~~~~~~~~~~~
\left(\frac{r\e}{r}\right)^{6}=27\,\frac{(1-\ep^2)(1-\es^2)}
{[1+(1-\ep^2)+(1-\es^2)]^3}\,.
\label{tecc2}
\eeq
To invert them, we must first solve the biquadratic equation
\beq
\es^4-3\left(1-\frac{r^2}{r\e^2}\frac{r\e^2}{a^2}\right)\es^2-\frac{r\e^2}{a^2}\left(3\frac{r^2}{r\e^2}-\frac{r\e^4}{a^4}\right)+2=0\,,
\label{invtecc1}
\eeq
and then replace the solution into the equation
\beq
\ep^2=3\left(1-\frac{r^2}{a^2}\right)-\es^2\,.
\label{invtecc2}
\eeq
Specifically, the only one solution of the biquadratic equation
(\ref{invtecc1}) guaranteeing the positiveness of the eccentricities
and the condition $\ep \ge \es$ is
\beq
\ep^2=\frac{3}{2}\left(1-\frac{r^2}{r\e^2}\frac{r\e^2}{a^2}\right)+\left[\frac{9}{4} \left(1-\frac{r^2}{r\e^2}\frac{r\e^2}{a^2}\right)^2+
\frac{r\e^2}{a^2}\left(3\frac{r^2}{r\e^2}-\frac{r\e^4}{a^4}\right)-2\right]^{1/2}
\label{invtecc21}
\eeq
\beq
\ep^2=\frac{3}{2}\left(1-\frac{r^2}{r\e^2}\frac{r\e^2}{a^2}\right)-\left[\frac{9}{4} \left(1-\frac{r^2}{r\e^2}\frac{r\e^2}{a^2}\right)^2+
\frac{r\e^2}{a^2}\left(3\frac{r^2}{r\e^2}-\frac{r\e^4}{a^4}\right)-2\right]^{1/2}\,.
\label{invtecc22}
\eeq
In the new variables, equations (\ref{6th}) and (\ref{8th}) take the
form (see the relations [\ref{tecc2}])
\beq
\frac{r\e(r)}{r}=\frac{r\e(r\col)}{r\col}~~~~~~~~~~~~{\rm and}~~~~~~~~~~~~
\frac{r}{r\e(r)}\frac{r\e^4(r)}{a^4(r)}=
\frac{1}{3}\left[1+\frac{5}{2}\,U(r)\left\lav \left(\frac{\delta\rho}{\srho}\right)^2\right\rav(r)\right]^{1/2}\left[\frac{r^6\e(r)}{a(r)^6}+3\frac{r^2}{r\e^2(r)}\frac{r\e^2(r)}{a^2(r)}-1\right]\,,
\eeq
which can be readily solved for the quantities $r\e(r)/r$ and
$r/a(r)$. Then, the eccentricities $\ep(r)$ and $\es(r)$ can be
obtained by means of equations (\ref{invtecc21}) and
(\ref{invtecc22}).

\end{document}